# Thermotunnel refrigerator with vacuum/insulator tunnel barrier: a theoretical analysis

Avto Tavkhelidze, Vasiko Svanidze, and Leri Tsakadze

Tbilisi State University, Chavchavadze ave. 13, 0179 Tbilisi, Georgia

E-mail: avtotav@geo.net.ge

We use two insulator layers in thermotunnel refrigerator, to modify the shape of the tunneling barrier so that electrons with high kinetic energy pass it with increased probability. Theoretical analysis show that overall tunneling current between the electrodes contains increased number of high kinetic energy electrons and reduced number of low energy ones, leading to high efficiency. Particular case of vacuum gap and solid insulator layer is calculated using digital methods. Efficiency remains high in the wide range of emitter electric field. Cooling coefficient is found to be as high as 40-50% in the wide range of emitter electric field.

Thermotunnel refrigeration at room temperature is widely discussed in literature [1-5]. Analysis of such systems shows that they have advantages over traditional thermoelectric refrigerators. Efficiency could be as high as 20-30 %. Besides it cooling power of 100 W/cm$^2$ could be achieved at room temperature[1]. Attempts of practical realization of such system underlined some problems [2, 5, 6]. Major problem is short circuit between the electrodes.

Thermotunnel refrigerator contains two conductive electrodes separated by vacuum gap of width of ~10 nm. Driven by applied electric field, electrons tunnel from emitter to collector carrying heat energy. Mirror forces and external electric field reduce the potential energy barrier. Tunneling electrons are distributed in wide range of initial kinetic energies. For efficient cooling it is necessary that electrons tunnel from high energy levels. We offer to coat collector with thin 3-5 nm layer of insulator. Insulator magnifies the electric field inside the vacuum gap. It changes the profile of the potential barrier so that high energy electrons tunnel trough barrier of less height and less width. At the same time low energy electrons tunnel through barrier of more width.

Potential of the electron between the conductive electrodes, with respect to image forces, has following form.

$$\begin{aligned} V(x) = {} & \Phi_1/q - V_1 x\,\theta(d_1-x)/d_1 - \theta(x-d_1)[(\Phi_1-\Phi_2)/q + V_1 + V_2(x-d_1)/d_2] + \\ & + (q^2/16\pi\varepsilon_0)\sum_{n=0}^{\infty}\sum_{i=0}^{n}\sum_{j=0}^{i} P_{n,i,j}\,k^i\,\{[\theta(x-d_1)/\varepsilon + \theta(d_1-x)][2(d+\alpha)^{-1} - (x+\alpha)^{-1} - (d-x+\alpha)^{-1}] + \\ & + k\theta(d_1-x)[(d_1-x+\alpha)^{-1} - 2(d_1+\alpha)^{-1} + (x+d_2+\alpha)^{-1}] + \\ & + k\theta(x-d_1)[2(d_2+\alpha)^{-1} - (x-d_1+\alpha)^{-1} - (2d-d_2-x+\alpha)^{-1}]/\varepsilon\} \end{aligned} \tag{1}$$

Where x – is distance from the emitter, $d_1$ – is thickness of vacuum gap between emitter and insulator layer, $d_2$ – is thickness of the insulator layer, $d=d_1+d_2$, q - is electron charge, $\Phi_1$ – work function of the emitter, $\Phi_2$ – work function of the collector, ε – high frequency dielectric constant of the insulator, $V_1= \varepsilon d_1 V_0/(d_2+\varepsilon d_1)$ – potential drop inside the vacuum gap, $V_2=d_2V_0/(d_2+\varepsilon d_1)$ – potential drop inside the dielectric layer, $V_0$ – is external voltage applied to the electrodes, $P_{n,i,j}=(-1)^{(i-j)}$ [n!/i!(n-i)!][i!/j!(i-j)!], $k=(1-\varepsilon)/(1+\varepsilon)$, $\alpha=(i-j)d_1+jd_2+(n-i)d$, and θ(x) – is step like function. Electron energy is given relative to Fermi energy of the emitter.

Fig. 1 shows potential profiles for the cases $d_2=0$, $d_2=30$Å and $d_2=50$Å. Potential barrier height and width reduce in the presence of insulator coating of the collector. Values of work functions and applied voltage are $\Phi_1=\Phi_2=1$eV, $V_0=1$V, ε=10 for all three profiles.

Integral tunneling current density contains electrons emitted from all energy levels up to potential barrier height – H

$$J_{tun}=\int_{-\infty}^{qH} N(E_x)D_{tun}(E_x)d\,E_x\,. \qquad (2)$$

Where $N(E_x)= mk_BT/2\pi^2\hbar^3\ln[1+\exp(-E_x/k_BT)]$ is number of electrons emitted from unit area during unit time having kinetic energy in the range of $[E_x, E_x+dE_x]$ and $D_{tun}(E_x)$ is the probability of tunneling of electron through potential barrier.

Fig. 1 shows that near the border of insulator and vacuum potential changes rapidly. Because of it, using the WKB method in that range will reduce the accuracy of calculation. Actually there is singularity in the potential profile at the border of insulator and vacuum. It is due to zero width of surface charge region. To avoid singularity we assume that mirror forces act only for distance that is more than some critical value. We choose critical value of 6 Å, because in practice the surface of insulator is not ideally plain but has roughness of the order of 5-10 Å. Formally we divide the distance between the emitter and collector in three parts. Vacuum gap, potential well and insulator layer. Inside the vacuum gap and insulator layer potential changes slowly and we use WKB approximation for tunneling probability calculations. Inside the potential well region we use formulas [3]:

$$D_2(E_x)=[1+H_0^2\sin^2(\sqrt{2m(E_x-H_0)}\,L/\hbar)/4E_x(E_x-H_0)]^{-1} \qquad H_0< Ex<H \qquad (3.1)$$

$$D_2(E_x)=[1+H_0^2\mathrm{sh}^2(\sqrt{2m(H_0-E_x)}\,L/\hbar)/4E_x(H_0- E_x)]^{-1} \qquad E_x<H_0 \quad , \qquad (3.2)$$

where $H_0$ is the depth of the well. In the near proximity of the potential well, we have very rapid change of potential due to image force singularity. It should be analyzed carefully because it corresponds to the infinite electric field, that is Physically impossible. We used following criteria to define the shape of the well. Surface roughness of the real insulator films is approximately 5 Å. Therefore, in reality, the border between two medias is defined within maximum 5 Å accuracy. We chose value of 6 Å for the well width because even number was convenient for interval separation. Besides it, we made calculations using 4 Å and 2 Å wide wells and confirmed that results do not change considerably.

Inside the vacuum gap, in the region x= 0 ÷ ($d_1$-4Å), and inside the dielectric in the region ($d_1$ +2Å) ÷ $d_2$ we use formula

$$D_{1,3}(E_x)=\exp[-2/\hbar \int_{x_1}^{x_2} \sqrt{2m(V(x)q-E_x)}\, dx] , \quad (4)$$

for tunneling probability calculation [1]. In Eq. (4), $x_1$ and $x_2$ are the solutions of equation $V(x)q-E_x=0$, where V(x) is inserted from Eq. (1).

Total tunneling probability will be the product of tunneling probabilities for each of three regions: $D_{tun}(E_x)= D_1(E_x)\, D_2(E_x)\, D_3(E_x)$. Heat density carried by tunneling electrons will be

$$Q_{tun} = \int_{-\infty}^{qH} (K_BT+ E_x)N(E_x)D_{tun}(E_x)dE_x \ . \quad (5)$$

Here $K_BT$ is mean kinetic energy of electron is x direction.

Electrons with energy $E_x$>H create thermionic current $J_R$ which we calculate using Richardson Equation. Heat density transferred by thermionic flow is $Q_R=(Hq+2k_BT)J_R$.

Fig.2 shows cooling power $Q_{tot}=Q_{tun}+Q_R$ as function of electric field applied to the emitter. Thicknesses of insulator layers are: $d_2$=30Å and $d_2$=50Å. This curves are compared to the curve $d_2$=0. All curves are calculated for $\Phi_1=\Phi_2$=1 eV, $V_0$=1V, $T_e$=300K. Fig. 2 shows that introducing of insulator layer reduces the cooling power.

Fig.3 shows dependence of useful cooling coefficient on applied field. Cooling coefficient was calculated as $\eta=Q_{tot}/ J_{tot}V_0$. Here $J_{tot}=J_{tun}+J_R$ is total current. Comparison of curves shows that introduction of insulator layer increase the cooling coefficient considerably. Cooling coefficient increase when thickness of the insulator layer increase. Besides it cooling coefficient becomes less field dependent, in the case of presence of the insulator.

Thermotunneling was realized using surface replication method, allowing preparation of conformal electrodes [7]. Second electrode was grown electrochemically on the first electrode and was conformal to it. Electrodes were prepared on the base of Si substrate. Thin films of Ti and Ag and Cs were grown on Si. Electrodes were separated using method or adhesion regulation between the thin films. Distance between electrodes was regulated using piezoelectric cylinder. Tunneling area of 1÷10 $mm^2$ was obtained in our experiments. To get considerable cooling effect it is necessary to introduce Ag-O-Cs electrodes [8], having work function 0.8 ÷1.1 eV. Ag-O-Cs electrodes are very reactive and require high vacuum environment. Coating of the collector electrode by insulator layer will help to solve the problem of work function stability. Another possibility of reduction of the electrode work function is utilizing of the quantum interference effect in thin metal films [9, 10]. It's main advantage is that it do not necessarily uses reactive materials like Cs.

There always will be heat backflow from hot to cold electrode. Heat conductivity of vacuum is formally zero. Backflow through thermal radiation [1] is less than 0.1 $W/cm^2$ at room temperature and is negligible. Backflow through housing depends on particular design and value as low as 0.1 $W/cm^2$ for $\Delta T$=30 K could be obtained in practice.

To summarize, we introduce insulator layer in the vacuum gap of thermotunnel refrigerator as coating of the collector. Insulator modifies potential profile in such manner that tunneling probability is increased for high energy electrons. As result cooling coefficient increases ~by 20-30% and reaches the value of 40-50%. At the same time cooling power reduce, but not dramatically. Cooling power of 1-3 W/cm$^2$, which is enough for most applications, is maintained for emitter electric fields of ~2 MV/cm with cooling coefficient of ~40 %. Both cooling power and cooling coefficient become less dependent on emitter electric field. In practice insulator layer will prevent electrodes from short circuit and damage. Fabrication of large area vacuum gap tunnel junction is quite problematic. Solving problem of short circuit makes possible new designs and will lead to increase of number of experiments in this direction.

The work was financed and supported by Borealis Technical Limited, assignee of corresponding patents (US 6,281,139; US 6,417,060; US 6,720,704; US 6,774,003; US 6,869,855; US 6,876,123; US 6,971,165).

## Figure Captions

Fig.1. Dependence of potential energy of electron on distance from emitter. (a) $d_2$=0, (b) $d_2$= 30 Å and (c) $d_2$= 50 Å. Barrier heights H and $H_0$ are shown for curve (c).

Fig.2. Dependence of heat flux between the electrodes on emitter electric field. (a) $d_2$=0, (b) $d_2$= 30 Å and (c) $d_2$= 50 Å.

Fig.3. Dependence of cooling coefficient on emitter electric field. (a) $d_2$=0, (b) $d_2$= 30 Å and (c) $d_2$= 50 Å.

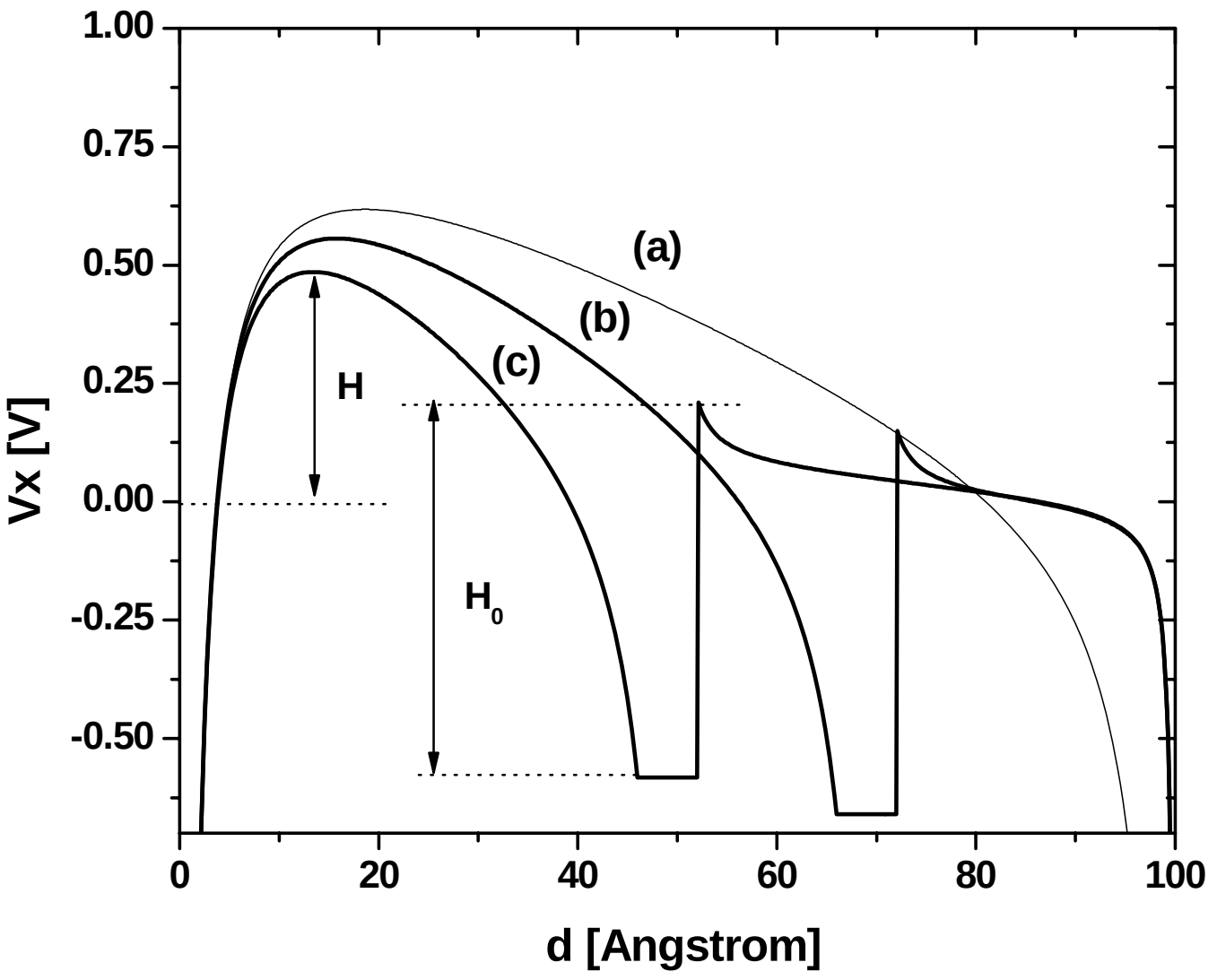

**Fig.1**

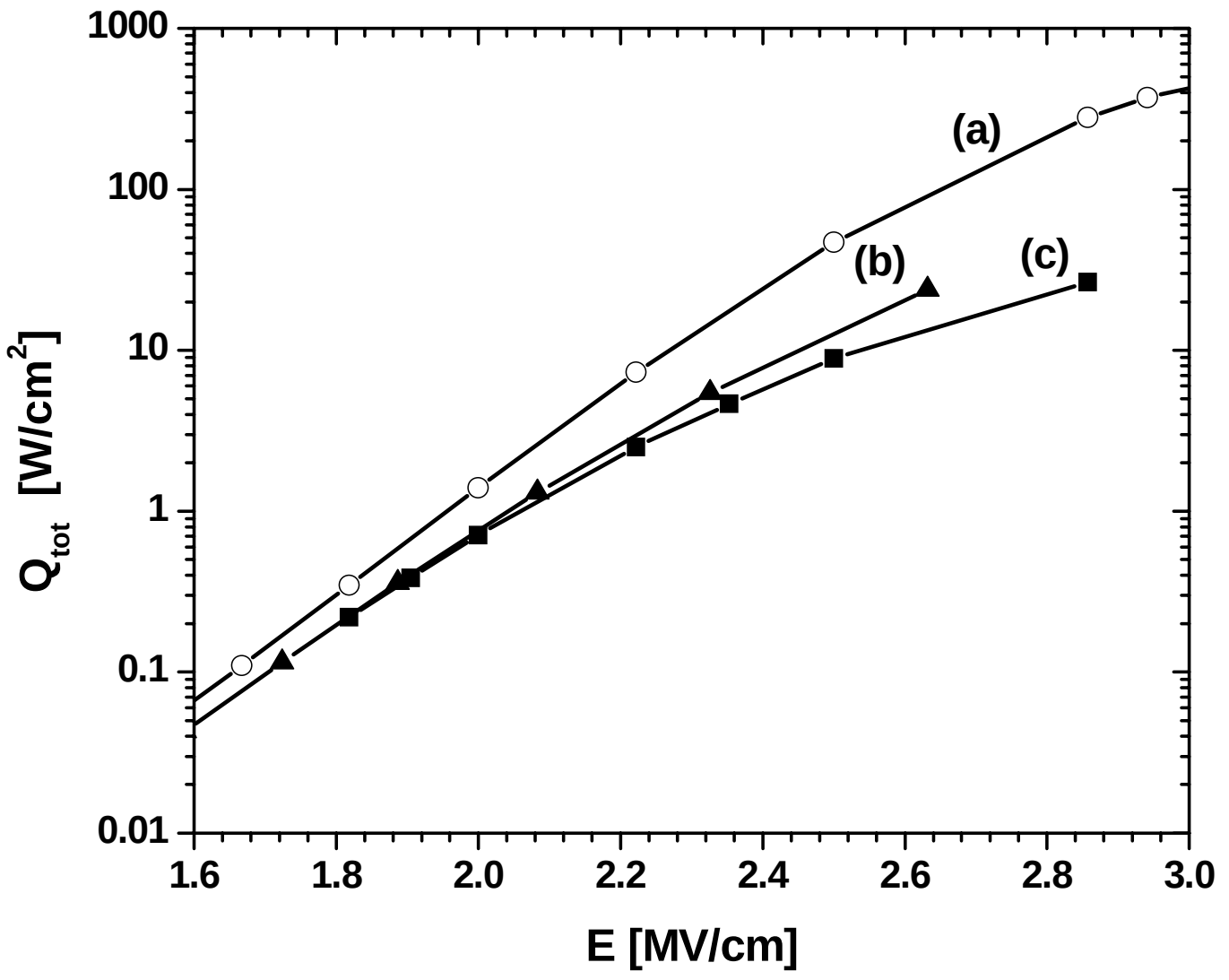

**Fig. 2**

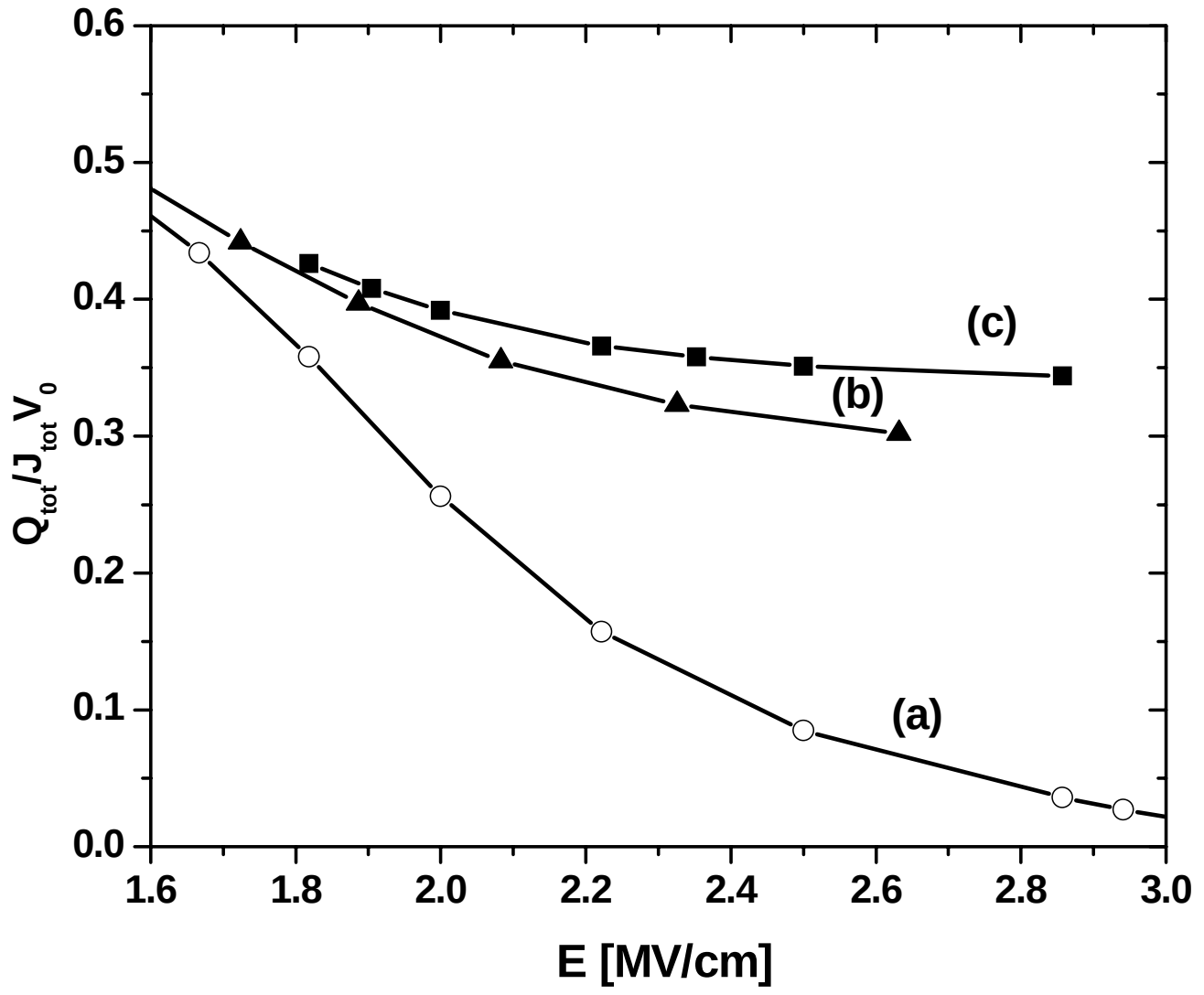

Fig. 3